\documentclass[%
 reprint,
 amsmath,amssymb,
 aps, 
 prl, 
 superscriptaddress, 
]{revtex4-2}

\usepackage{graphicx}% Include figure files
\usepackage{dcolumn}% Align table columns on decimal point
\usepackage{physics}
\usepackage{float} 
\usepackage{bm}% bold math
\usepackage{hyperref}% add hypertext capabilities
\usepackage{pdfpages}

\makeatletter
\AtBeginDocument{\let\LS@rot\@undefined}
\makeatother

\begin{document}

\preprint{APS/123-QED}

\title{Emergence of Interlayer Coherence in Twist-Controlled Graphene Double Layers}

\author{Kenneth A. Lin}
\affiliation{
  Microelectronics Research Center, Department of Electrical and Computer Engineering, The University of Texas at Austin, Austin, TX 78758, USA
}
\author{Nitin Prasad}

\affiliation{
  Department of Chemistry and Biochemistry, University of Maryland, College Park, MD 20742, USA
}
\author{G. William Burg}
\affiliation{
  Microelectronics Research Center, Department of Electrical and Computer Engineering, The University of Texas at Austin, Austin, TX 78758, USA
}
\author{Bo Zou}
\affiliation{
Department of Physics, The University of Texas at Austin, Austin, Texas 78712, USA
}

\author{Keiji Ueno}
\affiliation{
Department of Chemistry, Graduate School of Science and Engineering, Saitama University, Saitama 338-8570, Japan
}

\author{Kenji Watanabe}
\affiliation{
  Research Center for Functional Materials, National Institute of Materials Science, 1-1 Namiki Tsukuba, Ibaraki 305-0044, Japan
}
\author{Takashi Taniguchi}
\affiliation{
  International Center for Materials Nanoarchitectonics, National Institute of Materials Science, 1-1 Namiki Tsukuba, Ibaraki 305-0044, Japan
}
\author{Allan H. MacDonald}
\affiliation{
Department of Physics, The University of Texas at Austin, Austin, Texas 78712, USA
}
\author{Emanuel Tutuc}
\email[Corresponding Author: ]{etutuc@mail.utexas.edu}
\affiliation{
  Microelectronics Research Center, Department of Electrical and Computer Engineering, The University of Texas at Austin, Austin, TX 78758, USA
}

\date{\today}

\begin{abstract}
We report enhanced interlayer tunneling with reduced linewidth at zero interlayer bias in a twist-controlled double monolayer graphene heterostructure in the quantum Hall regime, when the top ($\nu_{\mathrm{T}}$) and bottom ($\nu_{\mathrm{B}}$) layer filling factors are near $\nu_{\mathrm{T}}=\pm1/2, \pm3/2$ and $\nu_{\mathrm{B}}=\pm1/2, \pm3/2$, and the total filling factor $\nu = \pm1$ or $\pm3$. The zero-bias interlayer conductance peaks are stable against variations of layer filling factor, and signal the emergence of interlayer phase coherence. Our results highlight twist control as a key attribute in revealing interlayer coherence using tunneling. 
\end{abstract}

\maketitle

In closely spaced double layer systems placed in the quantum Hall regime, the interlayer and intralayer interactions lead to ground states not present in single layers, including even denominator fractional quantum Hall states (QHSs) at total Landau level (LL) filling $\nu = 1/2$ and $1/4$ \cite{fracQHES1, fracQHES2, fracQHES3, fracQHES4}, as well as $\nu=1$ states that are interlayer electron-hole pair condensates \cite{excitonreview} with enhanced interlayer coherence. Experimental evidence for this phenomenon in GaAs double layers includes Josephson-like interlayer tunneling \cite{qhferro2, excitonGaAs4, PhysRevB.88.165308}, counterflow with near zero dissipation  \cite{excitonGaAs1, excitonGaAs2, excitonGaAs3, excitonGaAs5}, and Andreev reflection \cite{excitonGaAs6}. In graphene double layers, quantized Hall drag \cite{philipkim1} and counterflow \cite{crdeanExcitonDoubleBilayer} measurements have provided evidence of particle-hole pairing at total filling factor $\nu=1$ and $\nu=3$. Here, we investigate interlayer tunneling in a twist-controlled double monolayer graphene heterostructure, where tunneling in the quantum Hall effect regime provides insight into interlayer phase coherence. We observe enhanced interlayer tunneling at zero interlayer bias at $\nu=\pm1$ and $\pm3$ that is immune to changes in individual layer filling factors, a fingerprint of nascent interlayer phase coherence. 

\begin{figure}[H]
\center\includegraphics[width=0.4825\textwidth]{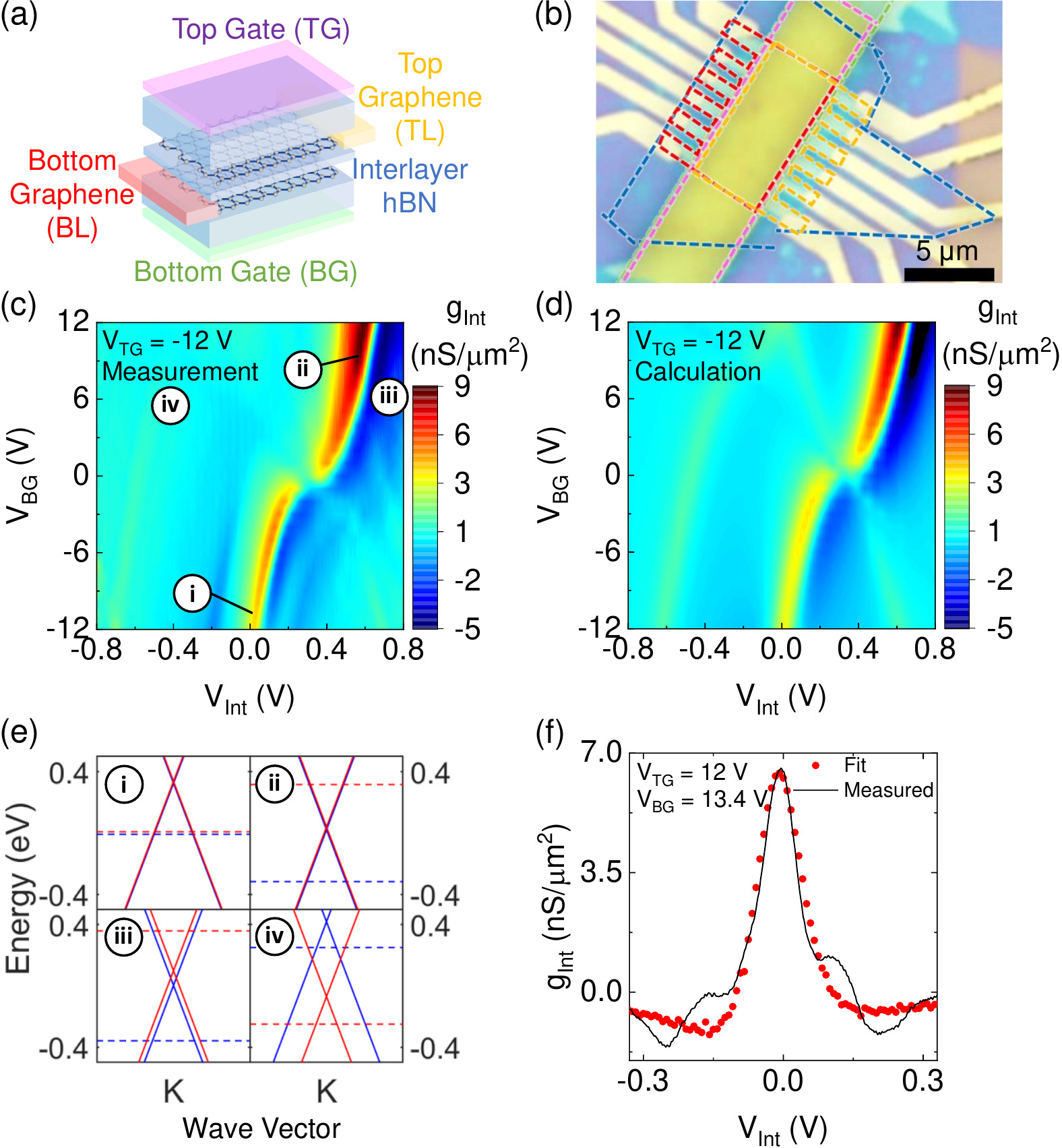}
\caption{\label{fig:fig1} (a) Schematic and (b) optical micrograph of a graphene double layer. (c) Experimental and (d) calculated $g_{\mathrm{Int}}$ vs. $V_{\mathrm{Int}}$ and $V_{\mathrm{BG}}$ at $T = 1.5$ K for $V_{\mathrm{TG}} = -12$ V. (e) Band alignment of top (red) and bottom (blue) graphene for the biasing points labeled in (c). The dashed lines indicate the layer Fermi levels. (f) Experimental (line) and calculated (dots) $g_{\mathrm{Int}}$ vs. $V_{\mathrm{Int}}$ at $V_{\mathrm{TG}} = 12$ and $V_{\mathrm{BG}} = 13.4$ V.}
\end{figure}

Figure \ref{fig:fig1}(a) shows a schematic of the twist-controlled double monolayer graphene-hBN heterostructure, which consists of two rotationally aligned and independently contacted graphene monolayers separated by a $d=2$ nm thick hBN tunnel barrier. Top and bottom hBN dielectrics encapsulate the heterostructure \cite{hBNflat}, and top ($V_{\mathrm{TG}}$) and bottom ($V_{\mathrm{BG}}$) graphite gate biases tune the layers densities. The crystal axis alignment of the two graphene layers \cite{highrotational} establishes energy and momentum conserving interlayer tunneling \cite{geimrestun, twistctrl, gatetunable, willsnanolett}, which leads to interlayer voltage-current characteristics with gate tunable negative differential resistance (NDR), and provides sensitivity to interlayer phase coherence. Multiple contacts to each graphene layer allow four-point interlayer current ($I_{\mathrm{Int}}$) vs. interlayer voltage  ($V_{\mathrm{Int}}$) measurements to decouple interlayer tunneling characteristics from contact resistances. Figure \ref{fig:fig1}(b) shows an optical micrograph of the heterostructure.

Figure \ref{fig:fig1}(c) shows the interlayer conductance $g_{\mathrm{Int}}=dI_{\mathrm{Int}}/{dV_{\mathrm{Int}}}$ vs. $V_{\mathrm{Int}}$ and $V_{\mathrm{BG}}$, measured at a top gate bias of $V_{\mathrm{TG}}=-12$ V, temperature $T=1.5$ K; $V_{\mathrm{Int}}$ is applied on the top layer, while the the bottom layer is held at ground. The data show resonant tunneling manifested by a $g_{\mathrm{Int}}$ peak and NDR, which evolve with $V_{\mathrm{Int}}$ and $V_{\mathrm{BG}}$, and correspond to the biasing condition where the layers energy bands are aligned, indicating energy and momentum conserving tunneling. We can explain the interlayer tunneling characteristics using a single-particle model \cite{willsnanolett, stronglyenhanced}, 
\begin{equation}
I_{\mathrm{Int}} = -e \displaystyle \int_{-\infty}^{\infty} T(E) \; \Big[ f_\mathrm{T}(E) -f_\mathrm{B}(E) \Big] \; dE
\end{equation}
where $E$ is the energy, $f_\mathrm{T}$ ($f_\mathrm{B}$) is the state occupancy in the top (bottom) layer, and $e$ the elementary charge. The tunneling rate [$T(E)$] is given by
\begin{equation}
T(E) = \dfrac{2\pi}{\hbar} \sum_{\vb{k}; s, s'} |t|^2 A_{\mathrm{T}}(\vb{k}, E) A_{\mathrm{B}}(\vb{k},E)
\end{equation}
where $t$ is the interlayer coupling, and $A_{\mathrm{T,B}}(\vb{k}, E)$ the spectral density in the top and bottom layers is assumed to be Lorentzian, 
\begin{equation}
A_{\mathrm{T,B}}(\vb{k}, E) = \dfrac{1}{\pi} \dfrac{\Gamma}{\left( E - \epsilon_{\mathrm{T,B}}(\vb{k}) \right)^2 +\Gamma^2}
\end{equation}
where $\epsilon_{\mathrm{T,B}}(\vb{k})$ is the top, bottom graphene energy-momentum dispersion, respectively, and $\Gamma$ the quasiparticle state energy broadening, assumed to be the same in both layers. The top (bottom) graphene layer density $n_\mathrm{T}$ ($n_\mathrm{B}$) is calculated using the following set of equations: 
\begin{eqnarray}
V_{\mathrm{BG}} && C_{\mathrm{BG}} +C{_{\mathrm{TG}}(V_{\mathrm{TG}} - V_{\mathrm{Int}})} = \nonumber\\ && e (n_{\mathrm{T}} + n_{\mathrm{B}}) + \dfrac{\mu_{\mathrm{B}} C_{\mathrm{BG}} + \mu_{\mathrm{T}} C_{\mathrm{TG}}}{e}
\end{eqnarray}
\begin{eqnarray}
V_{\mathrm{TG}} && C_{\mathrm{TG}} - V_{\mathrm{Int}} (C_{\mathrm{TG}} + C_{\mathrm{Int}}) = \nonumber\\ && e n_{\mathrm{T}} + \dfrac{\mu_{\mathrm{T}}}{e}(C_{\mathrm{Int}} + C_{\mathrm{TG}}) - \dfrac{\mu_{\mathrm{B}}}{e}C_{\mathrm{Int}} 
\end{eqnarray}
where $C_{\mathrm{TG}}$ ($C_\mathrm{BG}$) is the top (bottom) gate capacitance, $C_\mathrm{Int}$ the interlayer capacitance,  $\mu_{\mathrm{T}}$ ($\mu_{\mathrm{B}}$) is the top (bottom) layer chemical potential referenced to charge neutrality. 

Figure \ref{fig:fig1}(d) shows $g_{\mathrm{Int}}$ calculated  for the biasing conditions of Fig. \ref{fig:fig1}(c). An interlayer coupling of $t=1.0\;\mu$eV best fits the measurements.  Figure \ref{fig:fig1}(e) shows the calculated bands of the top and bottom graphene layers corresponding to the regimes labeled in Fig. \ref{fig:fig1}(c). At points (i) and (ii) a peak in interlayer conductance occurs because the energy bands are aligned. At points (iii) and (iv) the bands are energetically misaligned, suppressing the interlayer tunneling. Figure \ref{fig:fig1}(f) shows $g_{\mathrm{Int}}$ vs. $V_{\mathrm{Int}}$ measured at $V_{\mathrm{TG}}$ and $V_{\mathrm{BG}}$ values with the resonant peak at $V_{\mathrm{Int}}=0$ V. A fit of the experimental data yields $\Gamma=33$ meV. The $\Gamma$ values increase slightly with the layer density (see Fig. S1 of Supplemental Material). To assess the rotational alignment between the two layers, we performed calculations similar to Fig. \ref{fig:fig1}(d) data, but including a twist between the two layers. A comparison with experimental data indicate the alignment is within $0.2^\circ$ (see Fig. S2 and Fig. S3 of Supplemental Material).

\begin{figure}
\includegraphics[width=0.4825\textwidth]{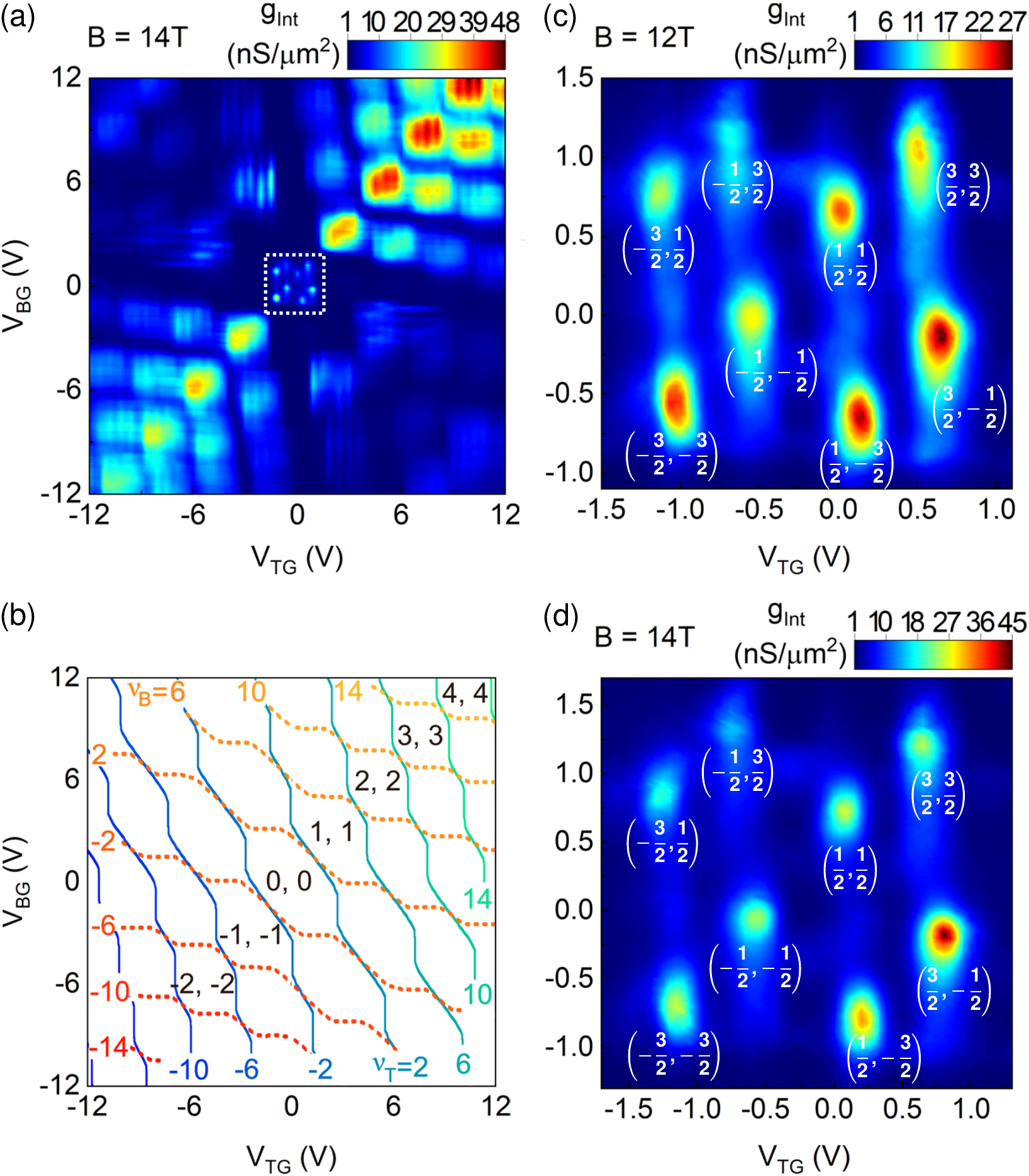}
\caption{\label{fig:fig2}(a) $g_{\mathrm{Int}}$ vs. $V_{\mathrm{TG}}$ and $V_{\mathrm{BG}}$ measured at $V_{\mathrm{Int}}=0$ V, $B=14$ T, and $T=1.5$ K. The dotted line marks the $N_\mathrm{T}=N_\mathrm{B}=0$ LL. (b) Calculated LL occupancy in each layer at $B=14$ T. The integers mark the orbital LL indices. (c)--(d) $g_{\mathrm{Int}}$ vs. $V_{\mathrm{TG}}$ and $V_{\mathrm{BG}}$ within the $N_\mathrm{T}=N_\mathrm{B}=0$ sector, at $B=12$ T [panel (c)] and $B=14$ T [panel (d)]. The $g_{\mathrm{Int}}$ maxima are labeled by their layer fillings $(\nu_T, \nu_B)$. }
\end{figure}

\begin{figure*} 
\includegraphics[width=0.99\textwidth]{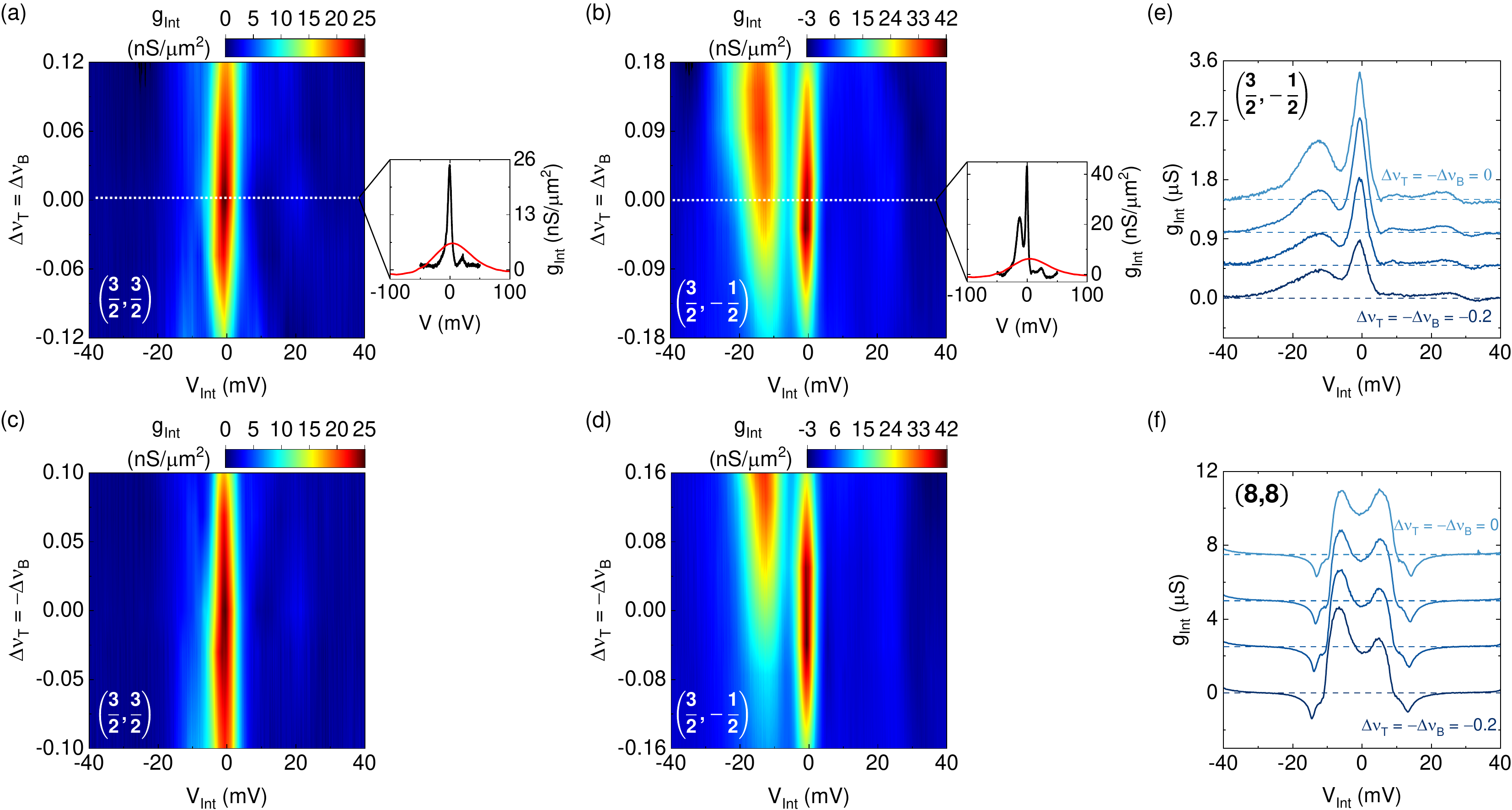}
\caption{\label{fig:fig3}(a)--(b) $g_{\mathrm{Int}}$ vs. $V_{\mathrm{Int}}$ and $\Delta \nu_{\mathrm{T}}=\Delta \nu_{\mathrm{B}}$ measured at $B=14$ T and $T=1.5$ K for (a) $(\nu_{\mathrm{T}}, \nu_{\mathrm{B}}) = (3/2,3/2)$, and (b) $(\nu_{\mathrm{T}}, \nu_{\mathrm{B}}) = (3/2,-1/2)$. The insets show $g_{\mathrm{Int}}$ vs. $V_{\mathrm{Int}}$ for $\Delta \nu_{\mathrm{T}}=\Delta \nu_{\mathrm{B}}=0$ at $B = 14$ T (black) and $g_{\mathrm{Int}}$ vs. interlayer electrostatic potential difference at $B = 0$ T (red). (c)--(d) $g_{\mathrm{Int}}$ vs. $V_{\mathrm{Int}}$ and $\Delta \nu_{\mathrm{T}}=-\Delta \nu_{\mathrm{B}}$ measured at $B=14$ T and $T=1.5$ K for (c) $(\nu_{\mathrm{T}}, \nu_{\mathrm{B}}) = (3/2,3/2)$, and (d) $(\nu_{\mathrm{T}}, \nu_{\mathrm{B}}) = (3/2,-1/2)$. (e)--(f) $g_{\mathrm{Int}}$ vs. $V_{\mathrm{Int}}$ at select $\Delta \nu_{\mathrm{T}}=-\Delta \nu_{\mathrm{B}}$ for (e) $(\nu_{\mathrm{T}}, \nu_{\mathrm{B}}) = (3/2,-1/2)$ and (f) $(\nu_{\mathrm{T}}, \nu_{\mathrm{B}}) = (8, 8)$. The traces are offset for clarity. The dashed lines mark $g_{\mathrm{Int}} = 0$.}

\end{figure*}

In a perpendicular magnetic field ($B$) the electrons occupy LLs, with a fourfold, spin and valley degeneracy in the absence of interactions \cite{grQH, grQH2}. In Fig. \ref{fig:fig2}(a), we plot $g_{\mathrm{Int}}$ measured at $V_{\mathrm{Int}}=0$ V as a function of $V_{\mathrm{TG}}$ and $V_{\mathrm{BG}}$ at $B=14$ T, and $T=1.5$ K (see Supplemental Material Fig. S4 for data measured at $B = 3$ T). The data show $g_{\mathrm{Int}}$ oscillations vs. $V_{\mathrm{TG}}$ and $V_{\mathrm{BG}}$, associated with LLs in both layers. To understand Fig. \ref{fig:fig2}(a) data, we employ Eqs. (4--5) and $\mu(N) = \mathrm{sgn}(N) v_F \sqrt{2 e \hbar B |N|}$, where $N$ is the highest occupied orbital LL index. Figure \ref{fig:fig2}(b) shows the  LL filling factor $\nu_{\mathrm{T}}$ ($\nu_{\mathrm{B}}$) in the top (bottom) layer, and the top (bottom) layer orbital LL indices $N_{\mathrm{T}} (N_{\mathrm{B}}$) along the $n_{\mathrm{T}} = n_{\mathrm{B}}$ diagonal. We determine $C_{\mathrm{TG}} = 88$ nF/cm$^2$, $C_{\mathrm{BG}} = 78$ nF/cm$^2$, and $C_{\mathrm{Int}} = 1.5$ $\mu$F/cm$^2$. The model accurately captures the experimental $g_{\mathrm{Int}}$ oscillations, with minima at $\nu_{\mathrm{T,B}} =\cdots,-10,-6,-2,2,6,10,\cdots$, consistent with a single-particle picture where $g_{\mathrm{Int}}$ minima (maxima) occur under full (partial) orbital LL fillings due to the availability of extended states. 

In Figs. \ref{fig:fig2}(c) and \ref{fig:fig2}(d), we highlight $g_{\mathrm{Int}}$ as a function of $V_{\mathrm{TG}}$ and $V_{\mathrm{BG}}$ with $V_{\mathrm{Int}}=0$ and $T=1.5$ K, inside the $N_{\mathrm{T}}=N_{\mathrm{B}}=0$ sector for $B=12$ and $B=14$ T respectively. Along the $n_{\mathrm{T}}=n_{\mathrm{B}}$ diagonal we observe clear $g_{\mathrm{Int}}$ maxima at $(\nu_{\mathrm{T}},\nu_{\mathrm{B}})=(\pm3/2,\pm3/2)$ and $(\pm1/2,\pm1/2)$, where both the top and bottom filling factors $\nu_{\mathrm{T}}$ and $\nu_{\mathrm{B}}$ are equal half-integers. The states ($\nu_{\mathrm{T}},\nu_{\mathrm{B}})=(\pm3/2,\pm3/2)$ correspond to a total filling factor of $\nu=\nu_{\mathrm{T}}+\nu_{\mathrm{B}}=\pm 3$, and ($\nu_{\mathrm{T}},\nu_{\mathrm{B}})=(\pm1/2,\pm1/2)$ correspond to $\nu=\pm 1$. In addition, we observe $g_{\mathrm{Int}}$ maxima at ($\nu_{\mathrm{T}},\nu_{\mathrm{B}})=(\pm3/2,\mp1/2)$ and $(\pm1/2,\mp3/2)$, corresponding to the imbalanced state at $\nu=\pm 1$. No peaks are observed at $(\nu_{\mathrm{T}},\nu_{\mathrm{B}})=(\pm3/2,\pm1/2)$ or $(\pm1/2,\pm3/2)$. Figures \ref{fig:fig2}(c) and 2(d) data depart markedly from observations made in $|N|>0$ LLs, where no $g_\mathrm{Int}$ peaks are observed when the layers are at half LL filling factors.

A mechanism that leads to enhanced $g_{\mathrm{Int}}$ values is the formation of interlayer phase coherent QHSs. Indeed, at $B=14$~T the effective layer separation ${d}/{l_B} =0.29$ is sufficiently small that inter- and intralayer interaction become comparable; $l_B=\sqrt{{\hbar}/{eB}}$ is the magnetic length. These conditions are expected to lead to phase coherence between electrons in different layers, which manifests in the case of short range order \cite{astern, Filsavchenko, timohyart} as an enhanced Josephson-like interlayer tunneling \cite{qhferro2, excitonGaAs4, PhysRevB.88.165308}.

To shed light on the mechanisms leading to enhanced tunneling at many half-integer layer fillings in Fig. \ref{fig:fig3}(a) we plot $g_{\mathrm{Int}}$ vs. $V_{\mathrm{Int}}$, when we concomitantly vary $\nu_{\mathrm{T}}$ and $\nu_{\mathrm{B}}$ by equal amounts $\Delta\nu_{\mathrm{T}}=\Delta\nu_{\mathrm{B}}$ away from $(\nu_{\mathrm{T}},\nu_{\mathrm{B}})=(3/2,3/2)$. The inset shows $g_{\mathrm{Int}}$ vs. $V_{\mathrm{Int}}$ when $\Delta\nu_{\mathrm{T}}=\Delta\nu_{\mathrm{B}}=0$ (black), and $g_{\mathrm{Int}}$ vs. interlayer electrostatic potential difference at $B = 0$ T (red), illustrating a much sharper zero-bias $g_{\mathrm{Int}}$ peak at $B = 14$ T compared to the $B=0$ T data. A similar dataset measured for $(\nu_{\mathrm{T}},\nu_{\mathrm{B}})=(3/2,-1/2)$ is shown in Fig. \ref{fig:fig3}(b). Figures \ref{fig:fig3}(c) and \ref{fig:fig3}(d) show $g_{\mathrm{Int}}$ vs. $V_{\mathrm{Int}}$ and $\Delta\nu_{\mathrm{T}}=-\Delta\nu_{\mathrm{B}}$ corresponding to $(\nu_{\mathrm{T}},\nu_{\mathrm{B}})=(3/2,3/2)$ and $(\nu_{\mathrm{T}},\nu_{\mathrm{B}})=(3/2,-1/2)$, respectively. Interestingly, for both $(\nu_{\mathrm{T}},\nu_{\mathrm{B}})=(3/2,3/2)$ and $(3/2,-1/2)$, corresponding to $\nu=3$  and $\nu=1$, $g_{\mathrm{Int}}$ peaks are observed at $V_{\mathrm{Int}}=0$ with widths significantly smaller compared to the $B=0$ T resonances [see e.g. Fig. \ref{fig:fig1}(f)]. The $g_{\mathrm{Int}}$ peak positions are stable at $V_{\mathrm{Int}}=0$ V, and do not respond to layer filling factor variations.
In contrast, the conductance in a noninteracting electron picture is proportional to an integral over energy of the layers density-of-states (DOS) product evaluated at $E$ in one layer and $E + eV_{\mathrm{Int}}$ in the other layer.  This picture predicts zero-bias $g_{\mathrm{Int}}$ peaks only when the DOS is maximized at the Fermi level in both layers, a property that cannot be maintained over finite ranges of layer filling factors. Indeed, calculations of $g_{\mathrm{Int}}$ vs. $V_{\mathrm{Int}}$ and $\Delta\nu_T = - \Delta\nu_B$ using a single-particle interlayer tunneling model for half-filled Landau levels (see Supplemental Material Fig. S5) show a $g_{\mathrm{Int}}$ peak that evolves with $V_{\mathrm{Int}}$, in clear contrast to Fig. \ref{fig:fig3}(a)--3(d) data.

The zero-bias $g_{\mathrm{Int}}$ peaks demonstrate the emergence of phase coherence between the two graphene monolayers at $\nu=1$ and $\nu=3$, where electrons occupy a coherent superposition of states in both layers. We contrast the zero-bias $g_{\mathrm{Int}}$ peaks observed at $\nu=1$ and $\nu=3$ with measurements at $\nu_{\mathrm{T}}=\nu_{\mathrm{B}}=8$ in the half-filled $N=2$ sector. Figures \ref{fig:fig3}(e) and \ref{fig:fig3}(f) compare $g_{\mathrm{Int}}$ vs. $V_{\mathrm{Int}}$ and $\Delta\nu_{\mathrm{T}}=-\Delta\nu_{\mathrm{B}}$ for $(\nu_T, \nu_B) = (3/2, -1/2)$ and $(8, 8)$ respectively. For the $(\nu_T, \nu_B) = (8, 8)$ case we observe $g_{\mathrm{Int}}$  minima at $V_{\mathrm{Int}}=0$, as a opposed to a peak, consistent with a tunneling gap at the Fermi level, similar to the suppression of tunneling between two half-filled uncorrelated LLs \cite{qhferro1, coulombgap}. We observe similar $g_{\mathrm{Int}}$ minima at $V_{\mathrm{Int}}=0$ at other half-filled LLs outside the $N=0$ sector, suggesting that interlayer phase coherence is present at $\nu=\pm1$ and $\nu=\pm3$, but not elsewhere.

\begin{figure} % was _4K_Fig4 with 0.78\textwidth
\includegraphics[width=0.4825\textwidth]{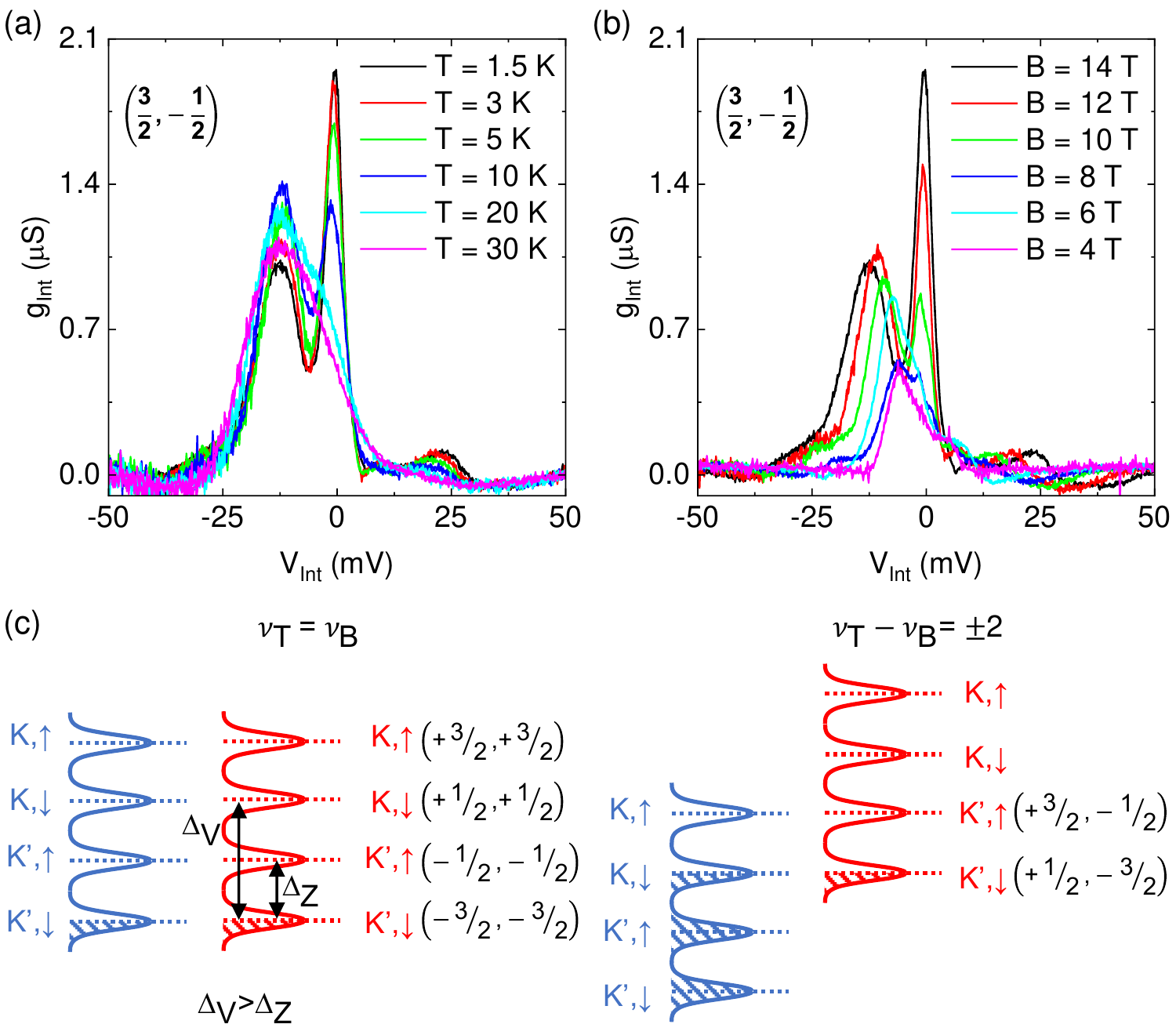}
\caption{\label{fig:fig4}(a) $T$ dependence of $g_{\mathrm{Int}}$ vs. $V_{\mathrm{Int}}$ at $B=14$ T, for $(\nu_{\mathrm{T}}, \nu_{\mathrm{B}}) = (3/2,-1/2)$. (b) $B$ dependence of $g_{\mathrm{Int}}$ vs. $V_{\mathrm{Int}}$ at $T=1.5$ K, for $(\nu_{\mathrm{T}}, \nu_{\mathrm{B}}) = (3/2,-1/2)$. (c) Schematic of the types of paired states observed. }
\end{figure}

The temperature dependence of the $g_{\mathrm{Int}}$ vs. $V_{\mathrm{Int}}$ zero-bias peak present at $(\nu_{\mathrm{T}},\nu_{\mathrm{B}})=(3/2,-1/2)$ is shown in Fig. \ref{fig:fig4}(a). The zero-bias $g_{\mathrm{Int}}$ peak for $(\nu_{\mathrm{T}},\nu_{\mathrm{B}})=(3/2,-1/2)$ decreases as $T$ is increased. A similar decrease in $g_{\mathrm{Int}}$ peak height as temperature is increased is observed at other $\nu=\pm1$ and $\pm3$ states (see Supplemental Material Fig. S6). At $T=30$ K, the interlayer conductance at $V_{\mathrm{Int}} = 0$ with the background tunneling removed ($\Delta g_{\mathrm{Int}}$) vanishes, signaling the interlayer phase coherence is no longer present. The $T$ dependence of $g_{\mathrm{Int}}$ is particularly interesting for $(\nu_{\mathrm{T}},\nu_{\mathrm{B}})=(3/2,-1/2)$, since the height of the zero-bias $g_{\mathrm{Int}}$ peak drops sharply with increasing temperature. In contrast, the $g_{\mathrm{Int}}$ side-peak at $V_{\mathrm{Int}}\neq0$ remains present and broadens noticeably with increasing temperature, which suggests the zero-bias peak is driven by interlayer phase coherence in the many-body ground state, whereas the side-peak is not. 
Figure \ref{fig:fig4}(b) shows the $B$ dependence of the $g_{\mathrm{Int}}$ vs. $V_{\mathrm{Int}}$ at $(\nu_{\mathrm{T}},\nu_{\mathrm{B}})=(3/2,-1/2)$. Reducing $B$ from $14$ T to $4$ T corresponds to varying $d/l_B$ from $0.29$ to $0.16$, which renders the double layer more interacting. However, the zero-bias $g_{\mathrm{Int}}$ peak for $(\nu_{\mathrm{T}},\nu_{\mathrm{B}})=(3/2,-1/2)$ decreases as $B$ is decreased, and is extinguished at $B = 6$ T, likely because of static disorder. 

Order in the interlayer electron-hole pair amplitude 
can be viewed as layer pseudospin ($\vec m$) ferromagnetism
with order in the $\hat{x}-\hat{y}$ plane. The layers' chemical potential difference is then $\propto m_z$, where $m_z$ is the pseudospin component along the $z$ axis.  According to the layer ferromagnet Landau-Lifshitz equations, it follows that nonequilibrium steady states with a fixed bias voltage are unstable to states with oscillatory collective dynamics. The enhanced tunneling seen experimentally in semiconductor quantum wells \cite{qhferro2, excitonGaAs4, PhysRevB.88.165308} have  been consistently interpreted as evidence for states with nascent order that has finite temporal and/or spatial range \cite{astern, Filsavchenko, timohyart}. The conductance due to enhanced short-range interlayer coherence is always peaked at zero bias, in contrast to single-particle resonant conductance peaks.

Because interlayer coherence is observed for $\nu_T = \nu_B$ and $\nu_T - \nu_B = \pm 2$, two types of paired states can be pictured [Fig. \ref{fig:fig4}(c)]. We assume that valley and spin degeneracy in the $N=0$ LL is lifted such that the valley splitting  ($\Delta_V$) dominates over the Zeeman effect ($\Delta_Z$) leading to a valley polarized $\nu=0$ QHS in each layer \cite{spinAndValleyQHFM}. The observation of interlayer coherence at both $\nu_T = \nu_B$ and $\nu_T - \nu_B = \pm 2$ is consistent with spin conservation in interlayer tunnneling. The absence of tunneling at $\nu_T - \nu_B = \pm 1,3$ is expected since single-particle spin-flip tunneling is expected to be extremely weak in graphene, and does imply that interlayer coherence is absent in these cases. If valley is also conserved in tunneling, the observation of interlayer coherence at $\nu_T - \nu_B = \pm 2$ implies that the valley splitting in the $N=0$ LL does not lead to $\bf{K}$ and $\bf{K'}$ states associated with the two sublattices of monolayer graphene, but rather a valley superposition \cite{scienceKekule,Coissard2022}.

Josephson-like interlayer tunneling associated with coherence in interacting double layers require single-particle tunneling because the critical current $I_0 \propto t^2$ \cite{astern, timohyart}. In twist-misaligned double layers single particle tunneling is suppressed because of the momentum mismatch between the band minima. The twist-aligned graphene double layer sample design ensures that single particle tunneling is not suppressed, and establishes twist control as key to probing interlayer coherence by identifying
tunneling anomalies in double layers of two-dimensional materials. 

\begin{acknowledgments}
We thank Timo Hyart for useful discussions. The work at The University of
Texas was supported by the National Science Foundation Grants No. EECS-2122476 and No. DMR-1720595, Army Research Office under Grant No. W911NF-17-1-0312. Work was partly done at the Texas Nanofabrication Facility supported by NSF Grant No. NNCI-1542159 and at the Texas Advanced Computing Center (TACC) at The University of Texas at Austin. K.W. and T.T. acknowledge support from the Elemental Strategy Initiative conducted by the MEXT, Japan (Grant No. JPMXP0112101001) and JSPS KAKENHI (Grant Nos. JP19H05790 and No. JP20H00354). K.U. acknowledges support from the JSPS KAKENHI Grants No. JP25107004, No. JP18H01822, No. JP21K04826, and No. JP22H05445.
\end{acknowledgments}

% NOTE: How to do the bibitem: in Overleaf, go to the Log view, and then at the lower right corner see Other Logs and Files, and select output.bbl, then paste it below. 

\bibliographystyle{apsrev4-2}
%\bibliography{refs}% Produces the bibliography via BibTeX.

\providecommand{\noopsort}[1]{}\providecommand{\singleletter}[1]{#1}%

\newpage
\includepdf[pages={1}]{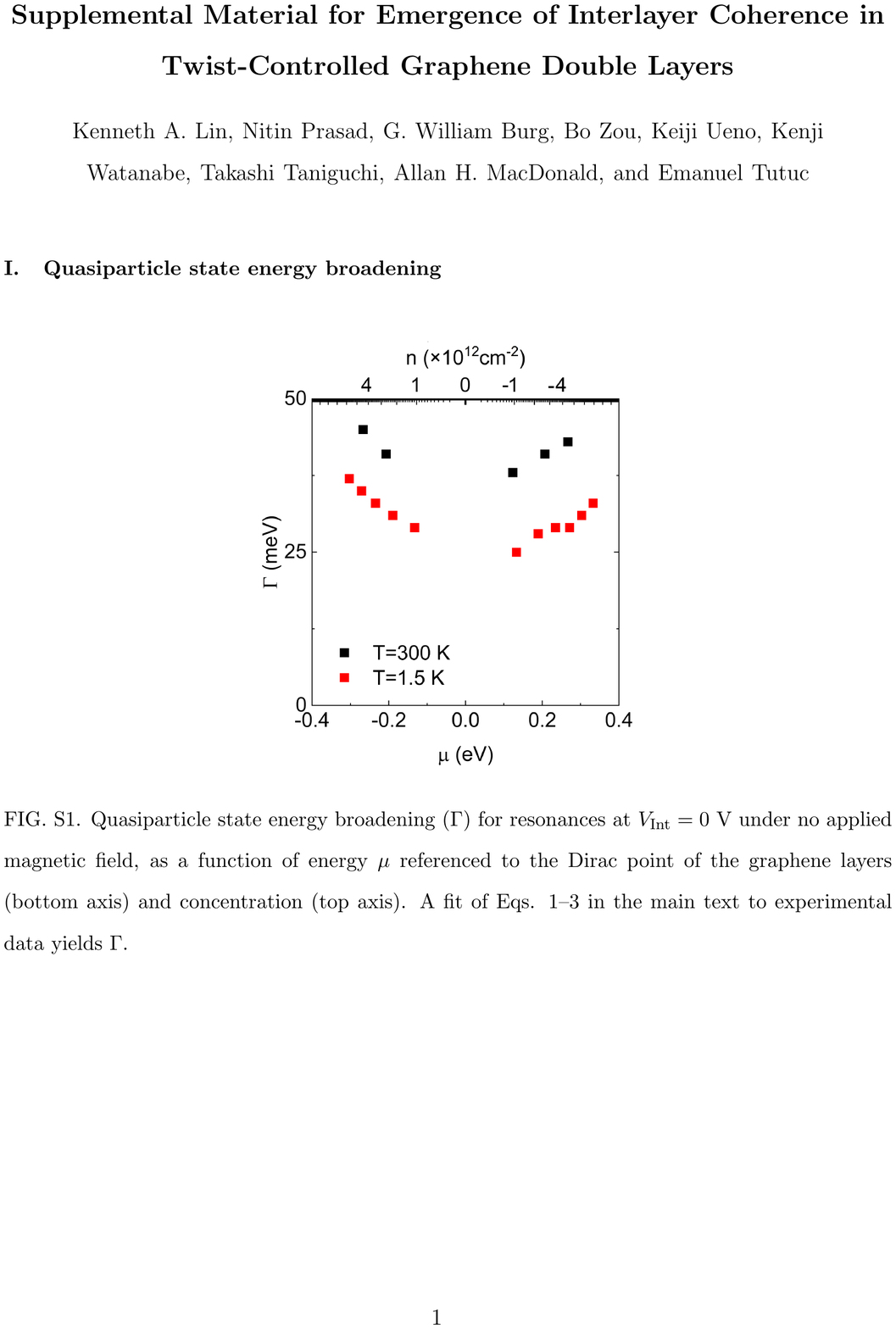}
{\color{white} .}
\newpage
\includepdf[pages={2}]{SM.pdf}
{\color{white} .}
\newpage
\includepdf[pages={3}]{SM.pdf}
{\color{white} .}
\newpage
\includepdf[pages={4}]{SM.pdf}
{\color{white} .}
\newpage
\includepdf[pages={5}]{SM.pdf}
{\color{white} .}
\newpage
\includepdf[pages={6}]{SM.pdf}
{\color{white} .}
\newpage
\includepdf[pages={7}]{SM.pdf}
{\color{white} .}
\newpage
\includepdf[pages={8}]{SM.pdf}

\end{document}